\documentclass[prl,twocolumn,superscriptaddress,showpacs]{revtex4}

\usepackage{graphicx}
\usepackage{dcolumn}
\usepackage{bm}

\begin{document}

\preprint{APS/123-QED}

\title{Nature of the Magnetic Order in the Charge-Ordered Cuprate La$_{1.48}$Nd$_{0.4}$Sr$_{0.12}$CuO$_4$}

\author{N. B. Christensen}
\affiliation{ Laboratory for Neutron Scattering, ETH Zurich
{\rm \&} Paul Scherrer Institute, CH-5232 Villigen PSI,
Switzerland }
\affiliation{ Materials Research Department, Ris{\o}
National Laboratory, Technical University of Denmark, DK-4000 Roskilde, Denmark }
\author{H. M. R{\o}nnow}
\affiliation{ Laboratory for Quantum Magnetism, {\'E}cole Polytechnique F{\'e}d{\'e}rale de Lausanne (EPFL), 1015 Lausanne, Switzerland }
\affiliation{ Laboratory for Neutron Scattering, ETH Zurich
{\rm \&} Paul Scherrer Institute, CH-5232 Villigen PSI,
Switzerland }
\author{J. Mesot}
\affiliation{ Laboratory for Neutron Scattering, ETH Zurich
{\rm \&} Paul Scherrer Institute, CH-5232 Villigen PSI,
Switzerland }
\author{R. A. Ewings}
\affiliation{ Department of Physics, Oxford University, Oxford, OX1
3PU, United Kingdom }
\author{N. Momono}
\affiliation{ Department of Physics, Hokkaido University, Sapporo 060-0810, Japan}
\author{M. Oda}
\affiliation{ Department of Physics, Hokkaido University, Sapporo 060-0810, Japan}
\author{M. Ido}
\affiliation{ Department of Physics, Hokkaido University, Sapporo 060-0810, Japan}
\author{M. Enderle}
\affiliation{ Institut Laue-Langevin, BP 156 - 38042 Grenoble Cedex 9 - France }
\author{D. F. McMorrow}
\affiliation{London Centre for Nanotechnology and Department of Physics and Astronomy,
University College London, UK}
\affiliation{ISIS Facility, Rutherford Appleton Laboratory, Chilton, Didcot, UK}
\author{A. T. Boothroyd}
\affiliation{ Department of Physics, Oxford University, Oxford, OX1
3PU, United Kingdom }

\begin{abstract}
Using polarized neutron scattering we establish that the magnetic
order in La$_{1.48}$Nd$_{0.4}$Sr$_{0.12}$CuO$_4$ is either (i)
one dimensionally modulated and collinear, consistent with the
stripe model or (ii) two dimensionally modulated with a novel
noncollinear structure. The measurements rule out a number of
alternative models characterized by 2D electronic order or 1D
helical spin order. The low-energy spin excitations are found to be
primarily transversely polarized relative to the stripe ordered state,
consistent with conventional spin waves.
\end{abstract}


\maketitle


One of the most striking and robust features in the phenomenology
of hole-doped copper oxide superconductors is the four-fold
incommensurate (IC) pattern of magnetic neutron scattering peaks
centered on the antiferromagnetic (AFM) wave vector of the square
CuO$_2$ lattice. This pattern is found in the magnetic excitation
spectrum of YBa$_2$Cu$_3$O$_{6+y}$ and
La$_{2-x}$(Sr,Ba)$_x$CuO$_4$ over wide doping ranges
\cite{incommensurate}. Near $x =1/8$ certain La-based materials
develop an elastic IC magnetic component accompanied by second
order harmonics around the structural Bragg peaks
\cite{Tranquada-Nature-1995,Tranquada-PRB-1996,LBSCO-stripes}. One
school of thought associates these features with one-dimensional
(1D) charge modulations separating AFM antiphase bands on the
CuO$_2$ layers \cite{stripes}. In this `stripe' model the
four-fold pattern is a superposition of two two-fold patterns,
arising from spatially separated stripe domains, each with charge
modulations along one of the two Cu--O bond directions. Static
stripes, posited to occur near $x =1/8$, are thought to compete
with superconductivity \cite{1/8-Suppression}, but dynamic stripes
could play a role in the formation of the superconducting state
\cite{Stripes-SC-theory}.

Recently, however, the stripe picture has been called into
question. Several new experimental findings point to the existence
of 2D charge density wave order in the ground state of hole-doped
cuprates \cite{STM,Komiya_PRL-2005}. In addition, the
dimensionality of the spin excitation spectrum is a subject of
debate \cite{spin-excitation-spectrum}. Furthermore, an increasing
number of phases exhibiting novel 2D electronic order have been
explored theoretically \cite{Sachdev-Science-2000}, including
orbital current correlations \cite{orbital-currents},
checkerboard-type orderings of Cooper pairs
\cite{checkerboard-pairs},
and 2D diagonal stripes \cite{Fine-PRB-2004}. The possibility that
features previously attributed to stripes might be
signatures of a more elaborate 2D ordering makes it vital to obtain
information with techniques that can separate different spatial
arrangements of spin and charge.

\begin{figure}[b]
\includegraphics[width=0.38\textwidth]{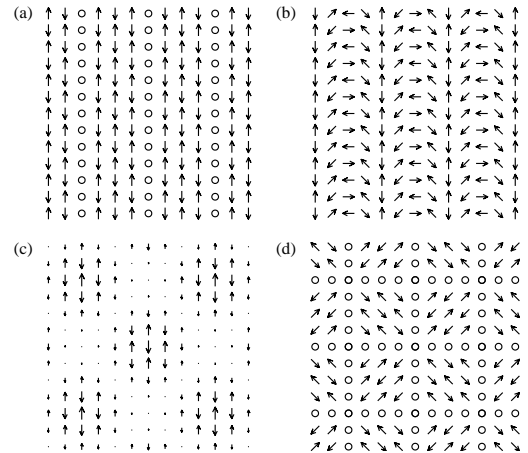}
\caption{\label{fig1} Models for the magnetic order in La-based
cuprates at $x=1/8$ doping. (a) One-${\bf q}$ domain with charge
stripes (lines of open circles) and collinear spin order
\cite{Tranquada-Nature-1995}. (b) One-${\bf q}$ domain with
helical spin order. (c) Collinear two-$\bf q$ structure of a
diagonally modulated commensurate AFM \cite{Fine-PRB-2004}. (d)
Two-$\bf q$ order of charge and spins in a noncollinear
structure.}
\end{figure}

Here we report a study of the magnetic order and dynamics in
La$_{1.6-x}$Nd$_{0.4}$Sr$_x$CuO$_4$ (LNSCO; $x=0.12$) by 
polarized-neutron scattering. Substitution of Nd for La stabilizes a
low-temperature tetragonal structure, which permits the formation,
below $\sim 50$\,K, of a robust spin--charge-ordered phase with
suppressed superconducting transition temperature $T_c$. The
relatively large ordered Cu moment $\sim 0.10$\,$\mu_B$
\cite{Tranquada-PRB-1996} allows detailed neutron polarisation
analysis. Below $T_{\rm Nd}\simeq 3$\,K, Nd--Cu coupling causes
alignment of the Nd spins along the $c$ axis with the same
ordering vectors as the Cu spins \cite{Tranquada-PRB-1996}. 
At temperatures well above $T_{\rm Nd}$ the Nd order is unlikely to influence 
the ordering pattern and in-plane direction of the Cu spins. 
We therefore believe that our study of static and dynamic
properties of LNSCO at $10$\,K has direct relevance to the
magnetic behaviour of Nd-free higher temperature superconductors.
Our results are most naturally understood in terms of a 1D
modulation of the AFM order, consistent with the occurrence of
stripes, although an exotic 2D noncollinear order is also
possible.

\begin{figure}
\includegraphics[width=0.420\textwidth]{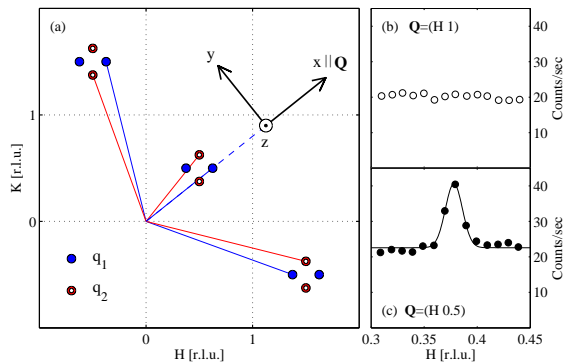}
\caption{\label{fig2} (a) Reciprocal space of the square CuO$_2$
lattice showing the three quartets of magnetic peaks investigated.
The peaks are displaced from AFM wave vectors by $\pm{\bf
q}_1=(\pm\delta,0)$ and $\pm{\bf q}_2=(0,\pm\delta)$ with $\delta
\simeq x = 0.12$. Also shown are the axes to which the
polarisation $\bf P$ and magnetic scattering are referred. (b)-(c)
Unpolarized elastic scans through the expected Bragg peak
positions for a $\pi$-spiral \cite{pi-spiral}.}
\end{figure}

Fig.\ \ref{fig1} shows four models yielding magnetic diffraction
patterns with principal Fourier components ${\bf Q}=(1/2, 1/2)\pm
{\bf q}_1$ and $(1/2, 1/2)\pm {\bf q}_2$ and equivalent
wave vectors, as found experimentally (Fig.\ \ref{fig2}). Figure\
\ref{fig1}(a) represents the conventional view
\cite{Tranquada-Nature-1995} that the quartet is due to incoherent
superposition of scattering from two equally populated domains
with collinear one-$\bf q$ spin order and orthogonal propagation
vectors. From the peak positions alone, this model is
indistinguishable from a model of two domains, each with helical
one-${\bf q}$ order \cite{Helical-Order} as sketched in Fig.
\ref{fig1}(b). The correct four-fold diffraction pattern is also
produced by the collinear two-$\bf q$ ``diagonal stripe" picture
\cite{Fine-PRB-2004} in Fig.\ \ref{fig1}(c) and by the
noncollinear two-$\bf q$ checkerboard structure shown in Fig.\
\ref{fig1}(d).


Our experiment was performed on the IN20 triple-axis spectrometer
at the ILL operated in Heusler-Heusler configuration with a pyrolytic graphite
filter to suppress higher order contamination of the scattered
beam. The crystal ($T_c=6.8$\,K), grown by the floating-zone
method, contained two grains separated by $\sim 1^{\circ}$. The sample
was mounted with the $[001]$ axis vertical in a He cryostat. A
final neutron energy of $34.8$ meV allowed access to IC quartets
surrounding several equivalent AFM wave vectors in the $(H,K,0)$
reciprocal lattice plane, denoted $(H,K)$ for short. The
polarisation vector ${\bf P}$ of the neutron beam at the sample
position was oriented along the $x$, $y$ and $z$ directions (see
below and Fig. \ref{fig2}). The scattered neutrons were recorded
in spin-flip (SF) and non-spin-flip (NSF) channels, according to
whether their spins had flipped or not on scattering. Corrections
for the measured beam polarisation $P = 0.86$ were applied. Most
measurements were performed at $10$\,K $>T_{c}>T_{\rm Nd}$, 
but some data taken at $1.7$\,K were used to fix the Bragg peak 
line shapes from the Nd magnetic order.


The cross section for scattering of polarized neutrons consists of
a purely nuclear term, a purely magnetic term and a
nuclear--magnetic interference term \cite{Moon1969}. We assume
that the latter can be neglected at the IC wave vectors of
interest, and that the nuclear spins are unpolarized. The magnetic
term contains two features that make it possible to determine
electronic spin directions. First, magnetic scattering originates
only from electronic spin components ${\bf S}_\perp$ perpendicular
to ${\bf Q}$. Second, SF scattering is caused by spin correlations
perpendicular to $\bf P$ \cite{Moon1969}. We denote coherent
nuclear scattering by $N$ and magnetic scattering by $M_x$, $M_y$
and $M_z$, where $M_{\alpha}$ is proportional to the time Fourier
transform of the correlation function $\langle S_{\alpha}(-{\bf
Q},0)S_{\alpha}({\bf Q},t)\rangle$ \cite{Squires}. We define axes
such that $x$ is parallel to $\bf Q$, and $y$ and $z$ are
perpendicular to $\bf Q$ in and out of the scattering plane,
respectively, (Fig.\ \ref{fig2}). With these axes, $M_x$ is
identically zero, and in the absence of a single-domain chiral
structure \cite{chiral} the intensities in the SF and NSF channels
with ${\bf P}$ parallel to $x$, $y$ and $z$ can be written
$I^x_{\rm SF}= M_y + M_z + B_{\rm SF}$, $I^y_{\rm SF}= M_z +
B_{\rm SF}$, $I^z_{\rm SF}= M_y + B_{\rm SF}$, $I^x_{\rm NSF}= N +
B_{\rm NSF}$, $I^y_{\rm NSF}= N +M_y + B_{\rm NSF}$ and $I^z_{\rm
NSF}= N +M_z + B_{\rm NSF}$, where $B_{\rm SF}$ and $B_{\rm NSF}$
are the backgrounds. These expressions apply both to elastic
($t=\infty$) and inelastic scattering.

We first treat elastic scattering, in which case $M_y$, $M_z$ are
proportional to the squares of the corresponding ordered spin
components. At each of the three IC quartets indicated in Fig.
\ref{fig2} the neutron count rate was recorded in some or all six
polarisation channels at each temperature. The purpose of probing
several zones is to vary the orientation of ${\bf Q}$ relative to
the ordered spin direction so that the magnetic cross sections
$M_y$ and $M_z$ change.

\begin{figure}[t!]
\includegraphics[width=0.40\textwidth]{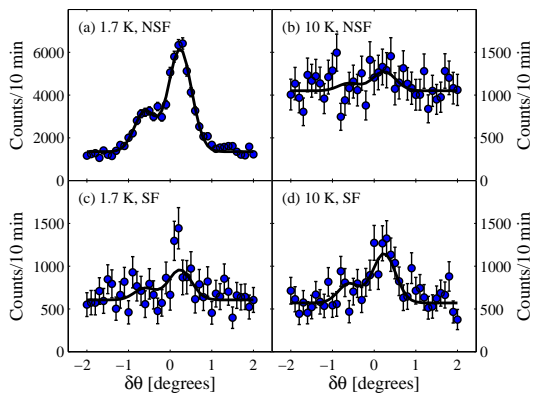}
\caption{\label{fig3} SF and NSF elastic scattering at $(1/2,
1/2)+{\bf q}_1$ with $\bf P \parallel$ {\it z}. The two-peak line
shape is due to the presence of two crystallites separated by $\sim
1^{\circ}$. The solid line in (a) is a fit to two Gaussians. The lines
in (b)--(d) were obtained from a fit to the same line shape as in (a)
but with the overall intensity scale and background allowed to
vary.}
\end{figure}


Fig.\ \ref{fig3} shows data obtained by scanning the sample
rotation angle $\theta$ through the $(1/2, 1/2)+{\bf q}_1$
satellite peak with ${\bf P} \parallel z$. The SF and NSF channels
then sense $M_y$ and $M_z$, respectively. The strong NSF signal at
$1.7$\,K [Fig. \ref{fig3}(a)] implies significant elastic
scattering from magnetic moments oriented perpendicular to the
CuO$_2$ planes. On heating to $10$\,K the NSF signal almost
vanishes [Fig. \ref{fig3}(b)]. A weak signal is present in the SF
channel at both $1.7$\,K and $10$\,K [Figs.\ \ref{fig3}(c) and
(d)]. These observations directly confirm an earlier finding that
the Nd moments order along the $c$ axis \cite{Tranquada-PRB-1996}.
Analysis of the data in Figs.\ \ref{fig3}(b) and (d) indicates
that at $10$\,K the spin direction is mainly confined to the
CuO$_2$ planes. The data do not rule out a small ordered component
along $c$, but as the presence or absence of such a component does
not alter our conclusions, which refer to the in-plane order, we
assume that at $10$\,K the Cu spins lie in the CuO$_2$ planes.

\begin{figure}[b!]
\includegraphics[width=0.40\textwidth]{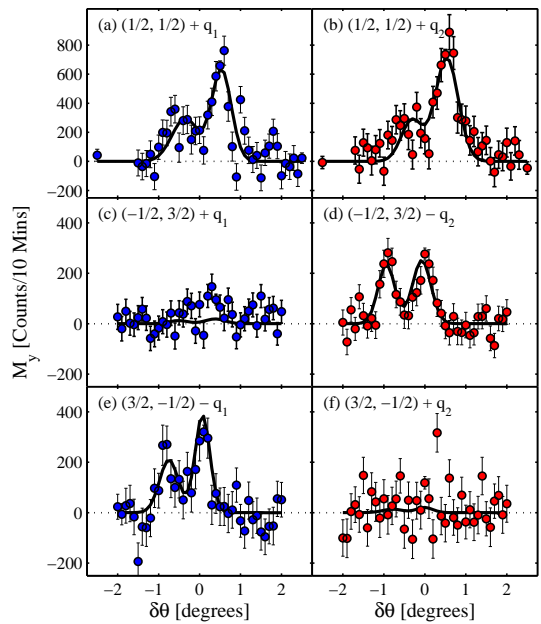}
\caption{\label{fig4} Processed polarized-neutron data taken at
$T=10$\,K, showing the in-plane component $M_y$ of the magnetic
cross section at each of six IC peaks of type ${\bf q}_1$ (left)
and ${\bf q}_2$ (right). $M_y$ was calculated both from the SF
data ($M_y=I^x_{\rm SF}-I^y_{\rm SF}$) and from the NSF data
($M_y=I^y_{\rm NSF}-I^x_{\rm NSF}$). After checking for
consistency, data thus obtained were combined. The solid lines
represent the model calculation described in the text.}
\end{figure}

We now turn to the spatial arrangement and in-plane orientation of
the Cu moments. Figure \ref{fig4} shows $M_y$ at three IC peaks of
type ${\bf q}_1$ and at three peaks of type ${\bf q}_2$. The ${\bf
q}_1$ and ${\bf q}_2$ peaks close to $(1/2, 1/2)$ have roughly equal
intensity. By contrast, near $(-1/2, 3/2)$ the ${\bf q}_1$ peak is
clearly weaker than the ${\bf q}_2$ peak, while the situation is
reversed near $(3/2, -1/2)$.

The relative intensities of the peaks in any given quartet depend
on (i) the orientation of ${\bf Q}$ relative to the spin
components contributing to the peaks and (ii) the population of
any equivalent magnetic domains. For the one-$\bf q$ order in
Fig.\ \ref{fig1}(b), the ${\bf q}_1$ satellites around $(-1/2,
3/2)$ should have same intensity as those around $(3/2, -1/2)$, in
disagreement with the data. This is true also for more than one
chiral domain \cite{Moon1969}, because all in-plane spin
directions contribute equally to each peak. Considering next the
collinear, two-$\bf q$ model in Fig.\ \ref{fig1}(c), the angle
between ${\bf Q}$ and the unique spin direction of a single domain
changes only slightly between ${\bf q}_1$ and ${\bf q}_2$ in any
given quartet, so the large intensity differences observed in
$M_y$ near $(-1/2, 3/2)$ and $(3/2, -1/2)$ cannot be reproduced by
this model. Hence, we can rule out both in-plane helical order
[Fig. \ref{fig1}(b)] and ``diagonal stripes" [Fig.\
\ref{fig1}(c)], as well as the so-called $\pi$-spiral model
\cite{pi-spiral} because that produces ${\bf q}_j$ satellites
about $(1/2, 1)$-type positions which are not observed, as shown
in Fig.\ \ref{fig2}(b) and \ref{fig2}(c).

By contrast, the data are consistent with two, equally populated,
one-$\bf q$ domains, each with collinear spin order, and spins
$\langle {\bf S} \rangle_j$ of domain $j$ approximately
perpendicular to ${\bf q}_j$ as shown for one domain in Fig.\
\ref{fig1}(a). In this scenario, the ${\bf q}_1$ and ${\bf q}_2$
peaks near $(1/2, 1/2)$ have nearly identical intensities because
the angle between ${{\bf Q}}$ and $\langle {\bf S} \rangle_j$ is
the same for both domains. Conversely, near $(-1/2, 3/2)$, the
${\bf q}_1$ (${\bf q}_2$) peak is relatively weak (strong) since
the corresponding spins are close to being parallel
(perpendicular) to ${\bf Q}$. Around $(3/2, -1/2)$ these angular
factors switch, and the ${\bf q}_1$ peak should be the most
intense, as observed. The solid lines in Fig.\ \ref{fig4} are
calculations based on this model. As in Fig.\ \ref{fig3}, the
line shape at each ${\bf Q}$ was fixed by fitting the large Nd
signal in $I^z_{\rm NSF}$ at $1.7$\,K. The corresponding curves
were then multiplied by the modulation expected for in-plane spins
$\langle {\bf S} \rangle_j$ perpendicular to ${\bf q}_j$ and
corrected for the form factor difference between the Nd-dominated
$1.7$\,K signal and the $10$\,K Cu signal. Although the
calculations do not agree in every detail, they clearly reproduce
the salient features of the data. Quantitatively, a least-squares
fit to all the data in Fig.\ \ref{fig4} results in $\langle {\bf
S} \rangle_1 \perp {\bf q}_1$ and $\langle {\bf S} \rangle_2 \perp
{\bf q}_2$ with a $\pm 3^\circ$ accuracy. Our diffraction data are
thus consistent with an {\it incoherent} superposition of orthogonal
stripe domains.

Our data are also consistent with the two-${\bf q}$ structure
shown in Fig.\ \ref{fig1}(d), which is a {\it coherent}
superposition of two orthogonal stripe domains and produces the
correct observed positions for both charge and magnetic peaks.
This has two implications. First, one cannot infer the existence
of 1D stripes from existing measurements of charge and magnetic
peak positions \cite{Tranquada-Nature-1995,stripes}, and second,
although our results do not rigorously rule out a checkerboard
configuration, they do impose the noncollinear spin arrangement
shown in Fig.\ \ref{fig1}(d) on any such model.


\begin{table}
\begin{ruledtabular}
\begin{tabular}{|lcr|}
Wavevector ${\bf Q}$ & Cross-section component & Intensity  \\
\hline
$(-1/2, 3/2)+{\bf q}_1$ & $M_y$ &  $7.4 \pm 1.3$ \\
$(-1/2, 3/2)+{\bf q}_1$ & $M_z$ &  $9.9 \pm 1.4$ \\
$(3/2, -1/2)-{\bf q}_1$ & $M_y$ &  $2.4 \pm 1.1$ \\
$(3/2, -1/2)-{\bf q}_1$ & $M_z$ &  $7.6 \pm 1.1$ \\
\multicolumn{2}{|c}{Excitation components relative to [010]} &  \\
\multicolumn{2}{|l}{Transverse,   out-of-plane} & $8.8 \pm 0.9$ \\
\multicolumn{2}{|l}{Transverse,   in-plane}     & $7.8 \pm 1.4$ \\
\multicolumn{2}{|l}{Longitudinal, in-plane}     & $1.7 \pm 1.3$ \\
\end{tabular}
\caption{\label{table1} Processed inelastic ($\hbar\omega=5$ meV)
count rates per 10 minutes at $T=10$\,K. $M_y$ and $M_z$ were
obtained from $I^x_{\rm SF}-I^y_{\rm SF}$ and $I^x_{\rm
SF}-I^z_{\rm SF}$, respectively. Total time: 44 h.}
\end{ruledtabular}
\end{table}


For inelastic scattering, $M_z$ is sensitive to spin fluctuations
out of the scattering plane and $M_y$ to in-plane fluctuations
perpendicular to ${\bf Q}$. Table \ref{table1} shows the count
rates at $(-1/2,3/2)+{\bf q}_1$ and $(3/2, -1/2)-{\bf q}_1$ at
$\hbar\omega=5$ meV. There is a statistically significant
difference between the values of $M_y$ at the two wave vectors.
This shows that the low-energy spin fluctuations in LNSCO have a
preferred direction, and it implies that they are not of the
singlet-triplet type (for which we would expect $M_y=M_z$). The
data are naturally explained by the ${\bf q}_1$ stripe domain
shown in Fig.\ \ref{fig1}(a). Converting from $M_y$ to components
transverse to and along $\langle {\bf S} \rangle_1$ shows (Table
\ref{table1}) that the fluctuations are predominantly transverse,
consistent with spin waves. This finding supports theories of the
cuprate spin excitation spectrum based on a ground state with
slowly fluctuating stripelike correlations, in which the 
low-energy excitations resemble Goldstone modes of weakly coupled spin
ladders \cite{Ladder-excitations}.


In summary, we have shown that the magnetic order in the
spin--charge ordered cuprate
La$_{1.48}$Nd$_{0.4}$Sr$_{0.12}$CuO$_4$ is either modulated in 1D
only (with spins perpendicular to the modulation direction) or
takes the form of a previously unconsidered noncollinear two-$\bf
q$ structure. In the former case, it is reasonable to conclude
that charge order is also one-dimensional, consistent with a
stripe model.


We thank J.P. Hill, J.M. Tranquada and S.A. Kivelson for stimulating
discussions. Support was provided by: Danish Natural Science Council
via DanScatt, Danish Technical Research Council Framework Program on
Superconductivity (N. B. C.), Wolfson Royal Society (D. F. M.), and EPSRC of Great Britain (R. A. E.).


\end{document}